\documentclass[3p,times,twocolumn]{elsarticle}
\usepackage{ecrc}

\volume{00}

\firstpage{1}

\journalname{Nuclear Physics B Proceedings Supplement}

\runauth{}

\jid{nuphbp}

\jnltitlelogo{Nuclear Physics B Proceedings Supplement}

\usepackage{amssymb,amsmath}
\usepackage[figuresright]{rotating}

\begin{document}

\begin{frontmatter}



\dochead{}

\title{Updates of PDFs for the 2nd LHC run}

\author[lbl1]{Patrick Motylinski\fnref{fn1}}
\author[lbl1]{Lucian Harland-Lang}
\author[lbl2]{Alan D.~Martin}
\author[lbl1]{Robert S.~Thorne}
%%\author[lbl2]{Ben Watt}

%\address{}
\address[lbl1]{Department of Physics and Astronomy, University College London, WC1E 6BT, UK}
\address[lbl2]{Institute for Particle Physics Phenomenology, Durham University, DH1 3LE, UK}

\fntext[fn1]{Speaker.}
%\author{}

%\address{}

\begin{abstract}
I present results on continuing updates in PDFs within the framework now called MMHT14 due to both theory improvements and the inclusion of new data sets, including most of the up-to-date LHC data. A new set of PDFs is essentially finalised, with no changes expected to the PDFs presented here.
\end{abstract}

%\begin{keyword}


%\end{keyword}

\end{frontmatter}

% \linenumbers


It has been more than five years since the publication of the global PDF analysis by MSTW titled `Parton distributions for the LHC`~\cite{MSTW}. Since then there have been several significant improvements in the data, in particular from the measurements made at the LHC, and it appears to be time for an new global PDF analysis within the previous MSTW08 (but now called MMHT14) framework.\\
As new data have become available they have been compared to the predictions provided by the MSTW PDFs. In the process we continued to use the extended parametrisation with Chebyshev polynomials as well as the freedom in deuteron nuclear corrections~\cite{MMSTWW}. This leads to a change in the $u_V-d_V$--distribution. Furthermore, we use the optimal GM-VFNS choice~\cite{Thorne} with its increased smoothing near the heavy flavour transition point. A small correction to the dimuon production ($\nu + N \rightarrow \mu^+ \mu^-$) has been taken into account for the case where the charm quark is produced away from the interaction point~\cite{Dimuon}. This has an impact, albeit rather small, on the strange distribution. Furthermore, issues regarding the charm branching fraction have been addressed. The value has been changed to $B_{\mu}=0.092 \pm 10\%$ from~\cite{Bolton} where the uncertainty is being fed into the PDFs.
In the MMHT framework we use the multiplicative definition of correlated uncertainties instead of additive~\cite{D'Agostini}. This way the uncorrelated errors effectively scale with the data.\\
\\
\section{Changes in data sets}
\label{}
We include new data that were officially published at the end of 2013.\\
There has been several additions of non-LHC sets, which should be mentioned here. The HERA run I neutral- and charged current data from both HERA and ZEUS have been replaced with a full combined set with treatment of the correlated errors~\cite{H1+ZEUS}. The HERA combined data on the $F_2^c(x,Q^2)$ structure function have been included~\cite{H1+ZEUScharm}. Furthermore published HERA data for $F_L(x,Q^2)$ measurements have been included as well~\cite{ZEUS-FL}. It has been decided to wait with the inclusion of Run II H1 and ZEUS until their combined data are available. \\
A whole range of Tevatron data sets have been included: 
CDF $W$--asymmetry data~\cite{CDF-Wasym}, D0 electron asymmetry data ($p_T>25$GeV, 0.75fb$^{-1}$)~\cite{D0-easym0.75} and D0 muon asymmetry data ($p_T>25$GeV, 7.3fb$^{-1}$)~\cite{D0-muasym7.3}. In addition, the final numbers for CDF $Z$-rapidity data have been taken into account~\cite{CDF-Zrap}. Overall, the inclusion of the mentioned sets do not have a major impact on the PDFs. The impact on $\alpha_S$ is rather moderate, too, with it changing to $\alpha_S=0.1199$ from  $\alpha_S=0.1202$ at NLO, and changing to $\alpha_S=0.1181$ from  $\alpha_S=0.1171$ at NNLO.\\

% % Correct dimuon cross-sections for missing small contribution, i.e. where charm is produced away from the interaction point. Previously assumed this was accounted for by acceptance corrections. Previous checks showed correction is a small effect on strange distribution.
%   \item Use \cb{NMC} structure function data with \cred{$F_L(x,Q^2)$}
% correction very close to theoretical \cred{$F_L(x,Q^2)$} value.   
% Very little effect.

%   \end{itemize}

\section{Inclusion of LHC data}
\label{}

\subsection{$W^{\pm}$, $Z$ and $t\bar{t}$}
\label{}

The inclusion of LHC data has been vastly facilitated by software such as \verb+FastNLO+, \verb+APPLGrid+, \verb+MCFM+, \verb+DYNNLO+ and \verb+FEWZ+. For instance, ATLAS $W^{\pm}$ and $Z$--rapidity data can be included directly in the fit~\cite{ATLAS-WZ}. Before this inclusion the global fit yielded $\chi^2\sim1.6$ per point at NLO and $\chi^2\sim 2$ per point at NNLO. After inclusion of the mentioned sets the value is brought down to $\chi^2\sim1.3$ at NLO, with the strongest pull being that on the gluon PDF. At NNLO the value is also brought down to $\chi^2\sim 1.3$, with the largest impact being that on the strange distribution.\\
% [, due to the balancing og $W$ and $Z$ is strongly dependent on the strange distribution].\\
\begin{figure}[t!]
  \centering
  \includegraphics[width=0.45\textwidth]{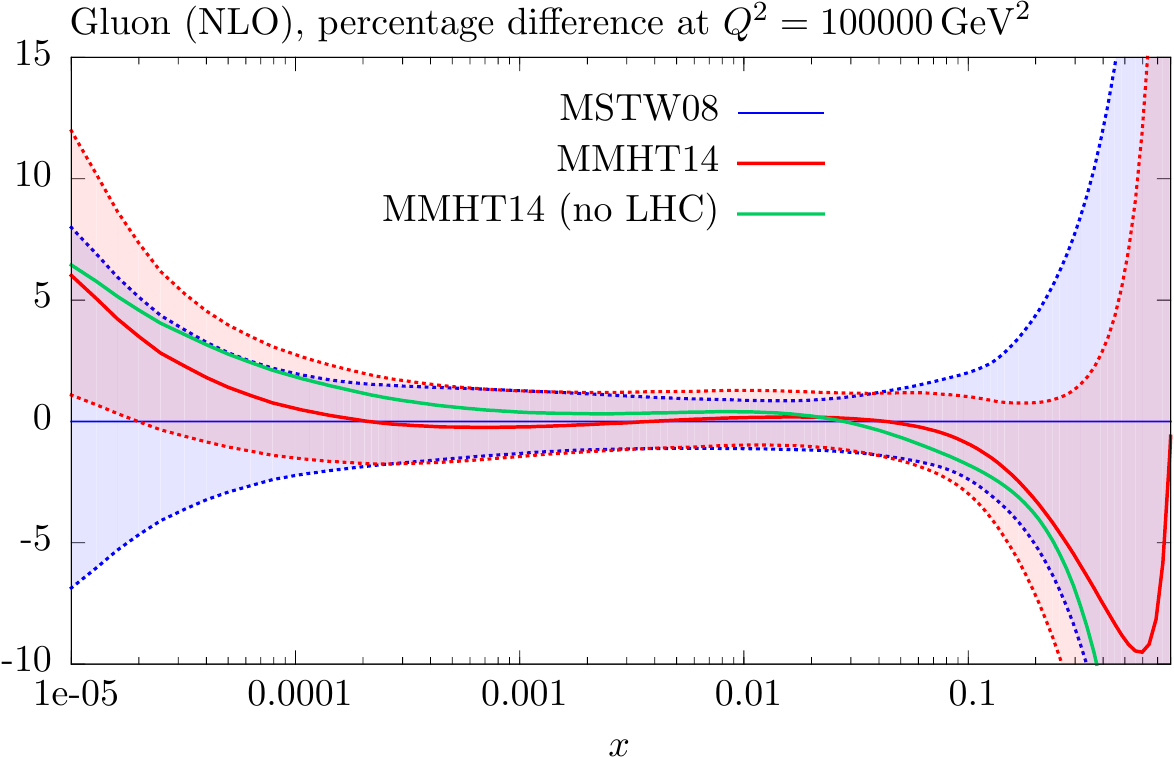}
  \includegraphics[width=0.45\textwidth]{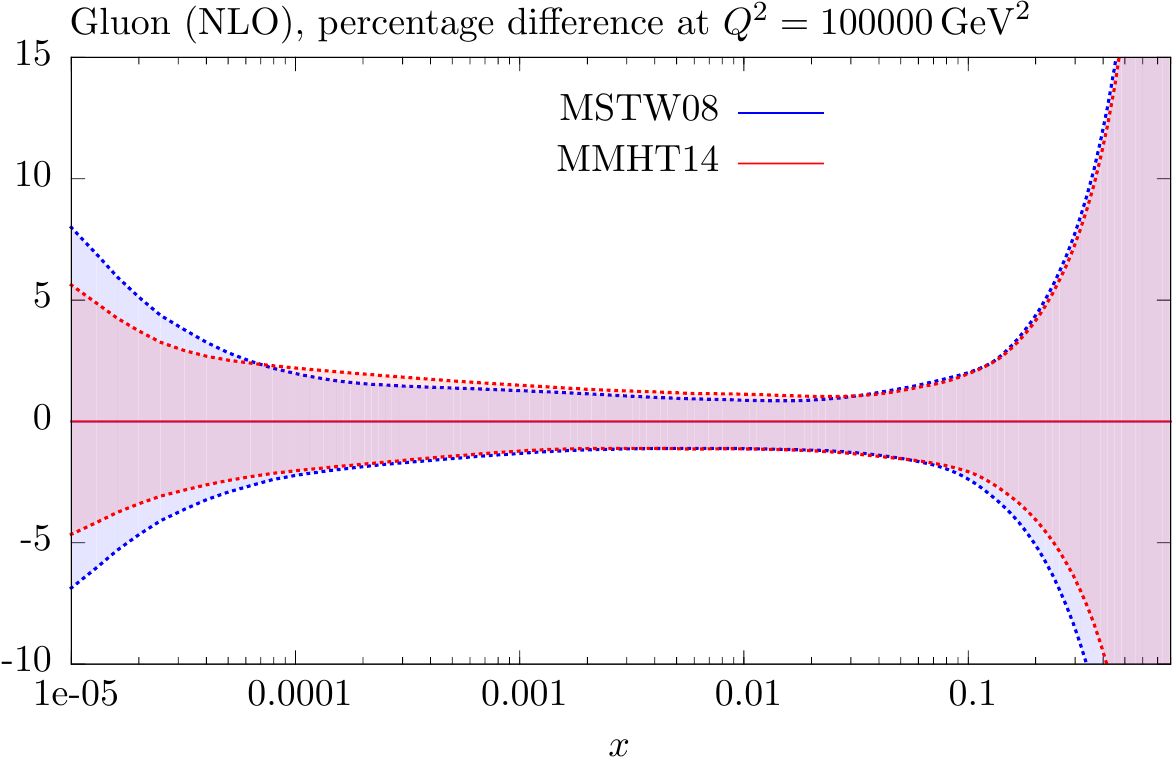}
  \caption{Upper pane: the change in NLO gluon distribution when including LHC data. Lower pane: the change in uncertainty when including LHC data. The central value for MMHT without LHC is also shown.}
  \label{fig:nloglu}
\end{figure}

\begin{figure}[t!]
  \centering
  \includegraphics[width=0.45\textwidth]{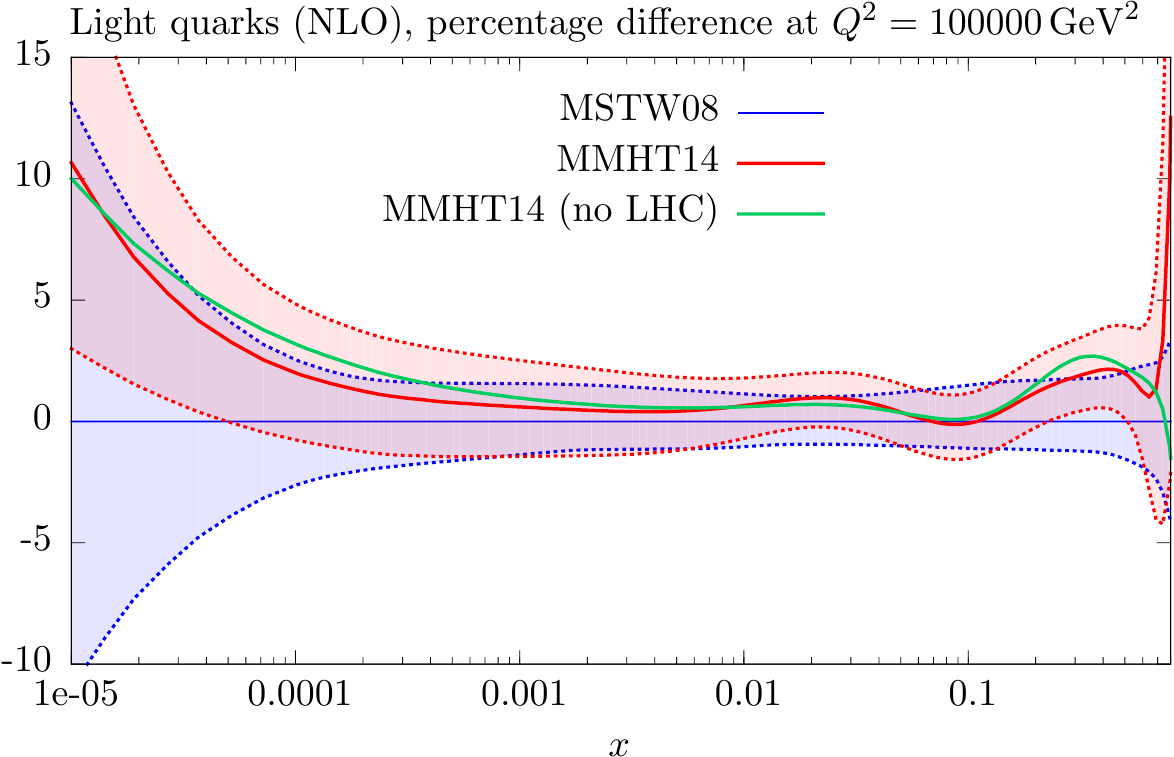}
  \includegraphics[width=0.45\textwidth]{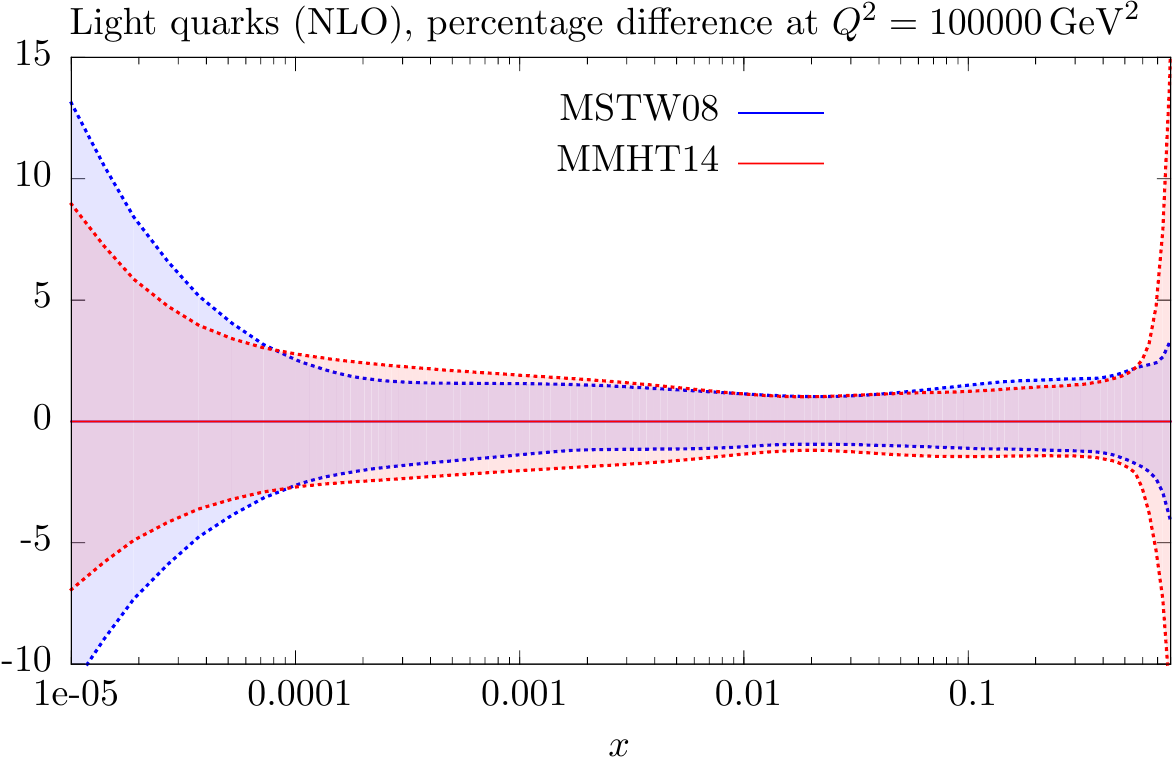}
  \caption{Upper pane: the change in NLO distribution of light flavour quarks when including LHC data. Lower pane: the change in uncertainty. The central value for MMHT without LHC is also shown.}
  \label{fig:lightnlo}
\end{figure}
Whereas MSTW08 fits showed a poor behaviour for the $u$--quark valence distribution at low $x$, this is no longer an issue with the new parametrisation used, and the inclusion of LHC data leads to further improvements. Apart from including  the ATLAS $W,Z$ already mentioned, in MMHT, we also include CMS data on $W$--asymmetry ($W$'s decaying into leptons)~\cite{CMS-Wasym,CMS-easym}, CMS $Z$--rapidity data~\cite{CMS-Zee}, LHCb data on $W^{\pm}$--production and $Z \rightarrow e^+ e^-$~\cite{LHCb-WZ,LHCb-Zee}, which all fit well at NLO. ATLAS high mass Drell-Yan data is also included and fits well, too.\\
More recently, we have also included the available data on $t\bar{t}$--production. More specifically this includes the combined cross section measurement from D0 and CDF as well as all published data from ATLAS and CMS at $\sqrt{S}=7$TeV and one point at $8$TeV. In doing so we have used $m_t=$172.5$\pm$1 GeV. Both NLO and NNLO fits behave well, with the NLO fit prefering masses slightly below 172.5 GeV and NNLO slightly above.\\
\mbox{}\\

\begin{figure}[t!]
  \centering
  \includegraphics[width=0.45\textwidth]{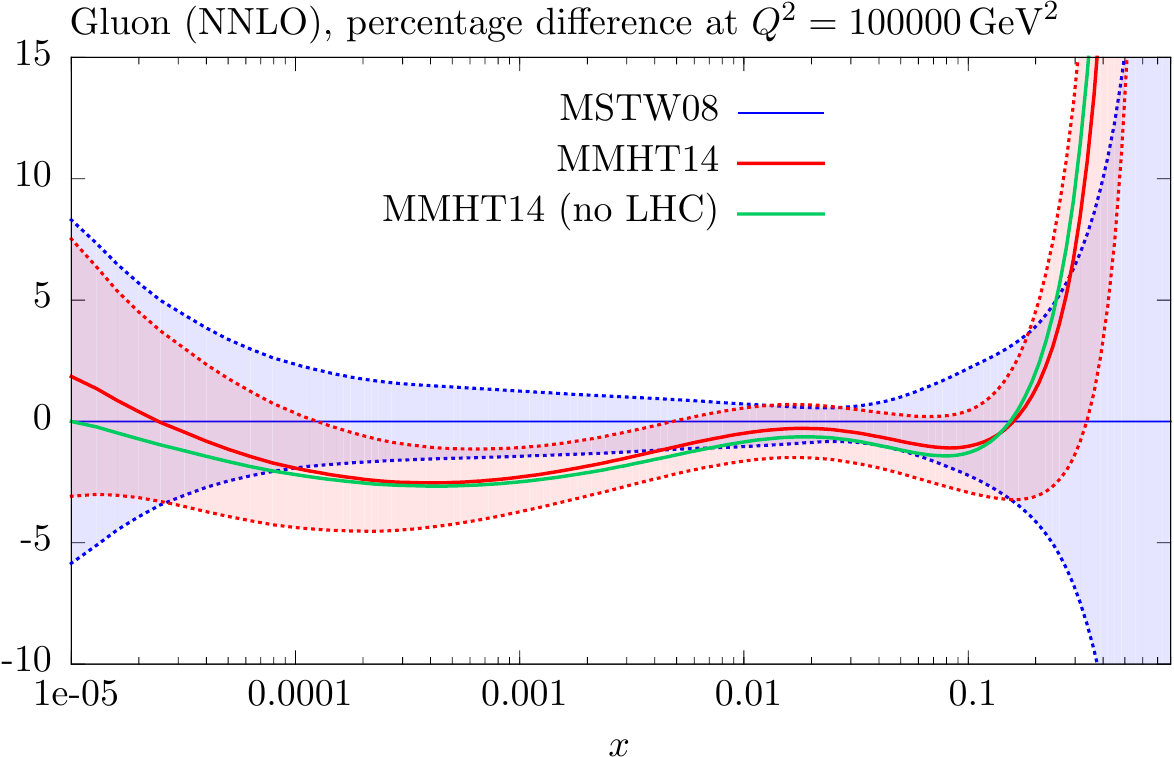}
  \includegraphics[width=0.45\textwidth]{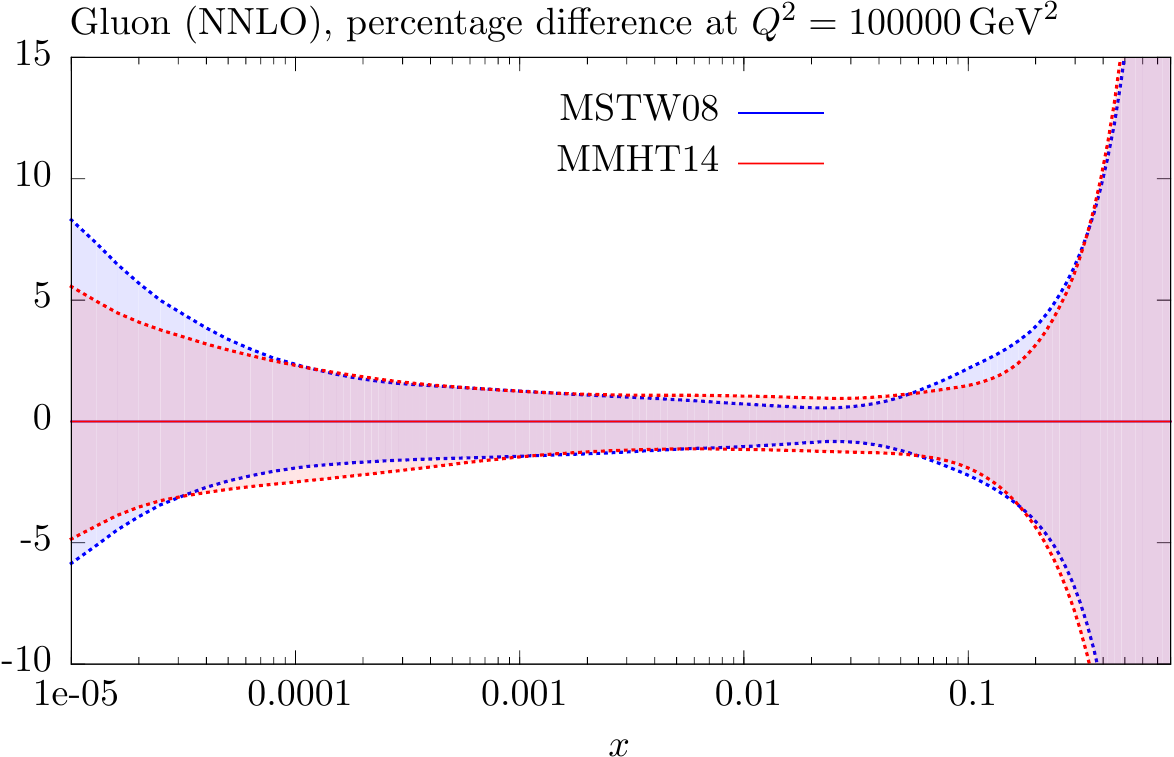}
  \caption{Upper pane: the change in NNLO gluon distribution when including LHC data. Lower pane: the change in uncertainty when including LHC data. The central value for MMHT without LHC is also shown.}
  \label{fig:nnloglu}
\end{figure}
\begin{figure}[t!]
  \centering
  \includegraphics[width=0.45\textwidth]{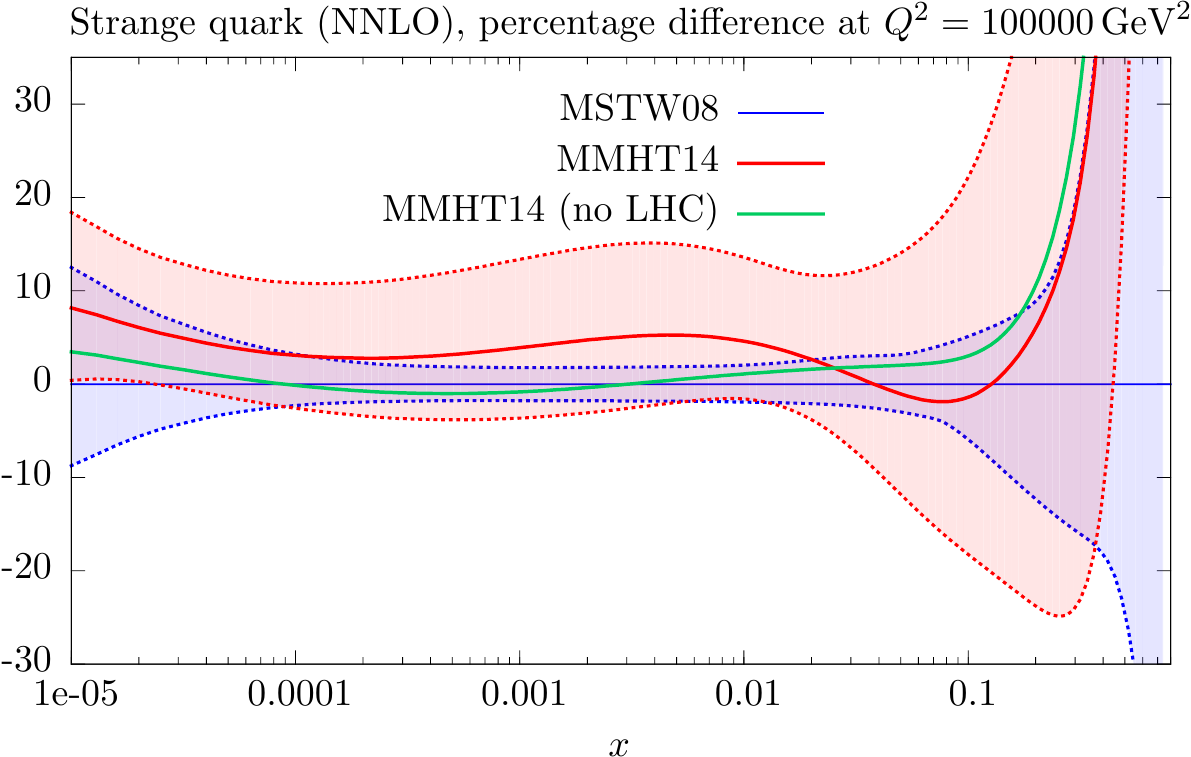}
  \includegraphics[width=0.45\textwidth]{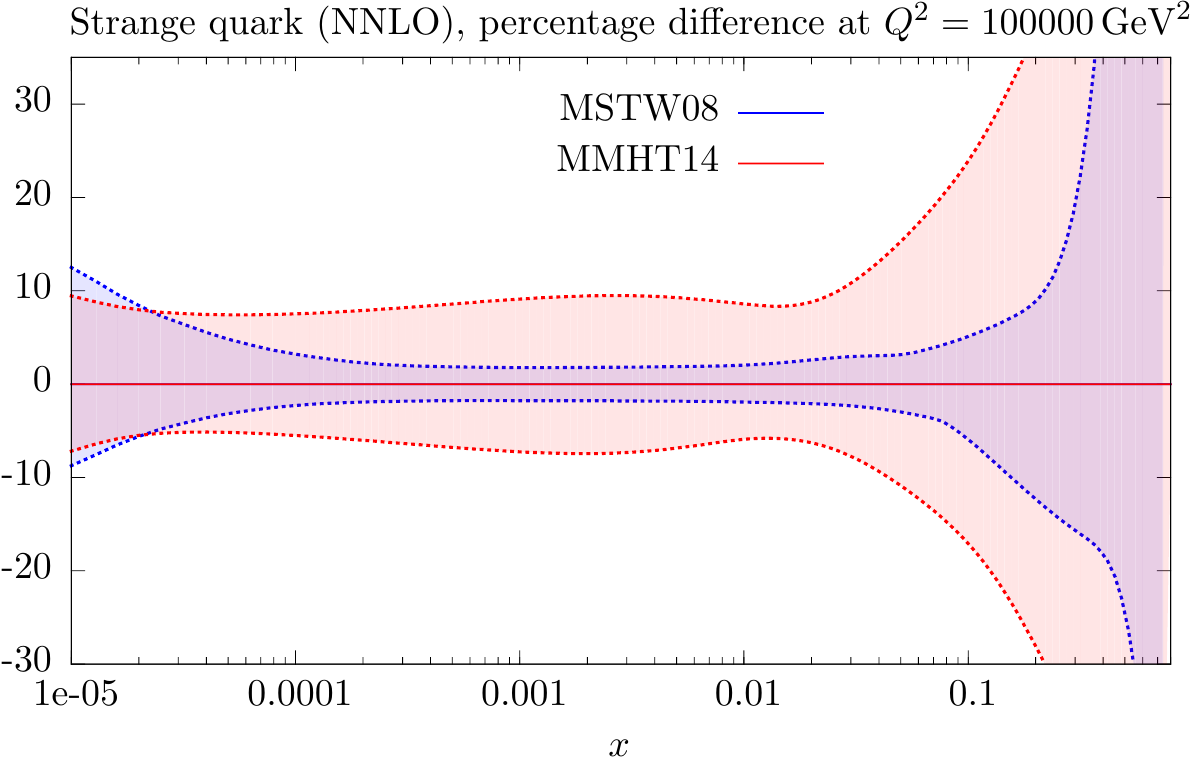}
  \caption{Upper pane: the change in NNLO gluon distribution when including LHC data. Lower pane: the change in uncertainty when including LHC data. The central value for MMHT without LHC is also shown.}
  \label{fig:strangennlo}
\end{figure}

\subsection{Jet data}
\label{}

At NLO CMS jet data at $\sqrt{S}=7$TeV~\cite{CMS-jet7} together with ATLAS $\sqrt{S}=7$TeV~\cite{ATLAS-jet7} and  $\sqrt{S}=2.76$TeV data~\cite{ATLAS-jet2.76} have been included in the fit. The simultaneous fit with both the CMS and ATLAS jet data sets lead to visible improvements, with the biggest improvement being that for the CMS data. \\
The CMS experiment has released an update to the jet data~\cite{CMS-jetREVISIT}. Previously the single pion uncertainties were published correlated, but recently a decision within the collaboration was made to decorrelate the single pion systematics. In practice this means the single pion source of uncertainty has been split into five parts. We have taken the decorrelated systematics into account in our fits. This leads to an improvement for these data with $\chi^2$ going from an initial 180 down to 140 and no major change in the PDFs.

\subsubsection{Jet data at NNLO}
\label{}

When considering inclusive jet production at NNLO in the context of PDF fits the situation is far less unanimous than at NLO. While the usage of threshold corrections in the context of Tevatron jet data is well justified~\cite{KO} as production there happens near threshold, in general, the situation is different for LHC data. Of course, for highish-$x$ it is similar at both experiments but the frequency at which production happens near threshold is much lower than at the Tevatron. The threshold corrections tend to increase significantly as we move away from threshold and, as a consequence, the quality of the fit deteriorates~\cite{WMT}.\\
The calculation of full fixed-order NNLO corrections for jet production are nearing completion~\cite{GGGP1,GGGP2}. Preliminary results already give some indications of the full form of the corrections. Near threshold the agreement with previous existing corrections is good but as we move away from threshold we can expect (positive) large corrections. In~\cite{THRESFIXED} threshold corrections have been compared to full fixed order calculations for the $gg$--channel at NNLO for both the Tevatron and LHC. While the discrepancy between threshold and fixed order corrections remains below a certain level at the Tevatron, usually not more than 30-50$\%$ at high rapidity then at the LHC the discrepancy reaches levels of several hundred percent at high rapidity and low $p_T$. This strongly illustrates the need for full fixed order corrections in the theory description.  \\
%Another issue, which has been addressed in the recent years is the $R$--dependence of the threshold corrections. At NLO this was investigated in~\cite{} where i was found that the threshold corrections vary considerably with $R$. More recently this has been addressed in the context of NNLO threshold corrections (\cite{}) where it has been shown that there is an additional $R$--dependence.\\
\\
In MMHT we have investigated how sensitive the NNLO fits are to varying the $K$--factors. In order to do so, we applied very approximate NNLO corrections (ad hoc $K$--factor parametrisations) with positive corrections of $\sim$5$\%$-20$\%$. The effects of applying a small and a large $K$--factor is show in Table~\ref{table:Kfac}. Neither of the $K$-factors change the PDFs much, with the smaller $K$--factor being the preferred. The changes are small compared to the uncertainty, and although the fit is slightly worse than at NLO it remains comparable. 
%This improves the fit somewhat but the improvement appears to be more linked to the previously mentioned change in the treatment of the CMS jet data systematics. 

\begin{table*}[t]
\begin{center}
%\vspace{-.5cm}
\begin{tabular}{|l|c|c|c|c|}
\hline
 &  & MMSTWW     & MMHT2014 & MMHT2014 \\
~~~~~~data set         &   {$N_{pts}$}          & Ref.\cite{MMSTWW}        & (no LHC)       &  (with LHC)       \\
\hline
NLO  \\
\hline
ATLAS jets ({2.76~TeV+7~TeV})        & 116& 107 & 107 & 106 \\   
CMS jets ({7~TeV})                   & 133& 140 & 143 & 138 \\   
\hline
NNLO small $K$-factor \\
\hline
ATLAS jets ({2.76~TeV+7~TeV})        & 116& (107) & (123) & (119) 115\\   
CMS jets ({7~TeV})                   & 133& (142) & (137) & (135) 137\\   
\hline
NNLO large $K$-factor \\
\hline
ATLAS jets ({2.76~TeV+7~TeV})        & 116& (117) & (132) & (128) 126\\   
CMS jets ({7~TeV})                   & 133& (145) & (137) & (139) 139\\   
\hline
    \end{tabular}
\end{center}
%\vspace{-.0cm}
\caption{\sf The quality of the description (as measured by the value of 
$\chi^2$) of the LHC inclusive jet data both before and after 
they have been included in the global NLO and NNLO fits. 
For comparison we also show the $\chi^2$ values obtained in 
the CPdeut fit of the NLO MMSTWW analysis~\cite{MMSTWW}, which did not 
include LHC data. The bracketed numbers are for the cases where the LHC jet data was not included in the NNLO fit.}
%\label{tab:LHCjet}
  \label{table:Kfac}
\end{table*}

\begin{table*}[t]
\begin{center}
%\vspace{-.5cm}
\begin{tabular}{|l|c|c|c|c|}
\hline
 &  & MMSTWW     & MMHT2014 & MMHT2014 \\
~~~~~~data set         &   {$N_{pts}$}          & Ref.\cite{MMSTWW}        & (no LHC)       &  (with LHC)       \\
\hline
NLO  \\
\hline
ATLAS {$W^+, W^-, Z$}               & 30 & 47 & 44 & 38 \\   
CMS {$W$} asymm {$p_T >35~$GeV}    & 11 & 9  & 16 & 7  \\    
CMS asymm {$p_T >25~$GeV,$30~$GeV}  & 24 & 9  & 17 & 8  \\    
LHCb {$Z\to e^+e^-$}                & 9  & 13 & 13 & 13 \\   
LHCb {$W$} asymm {$p_T >20~$GeV}   & 10 & 12 & 14 & 12 \\
CMS  {$Z\to e^+e^-$}                & 35 & 21 & 22 & 19 \\
ATLAS high-mass Drell-Yan           & 13 & 20 & 20 & 21 \\
CMS double diff. Drell-Yan          & 132& 385  & 396  & 373  \\   
\hline
NNLO\\
\hline
ATLAS {$W^+, W^-, Z$}              & 30 & 72 & 53 & 39  \\ 
CMS {$W$} asymm {$p_T >35~$GeV}   & 11 & 18 & 15 & 8 \\    
CMS asymm {$p_T >25~,30~$GeV}     & 24 & 18 & 17 & 9 \\    
LHCb {$Z\to e^+e^-$}               & 9  & 23 & 22 & 21 \\   
LHCb {$W$} asymm {$p_T >20~$GeV}  & 10 & 24 & 21 & 18 \\
CMS  {$Z\to e^+e^-$}               & 35 & 30 & 24 & 22 \\
ATLAS high-mass Drell-Yan          & 13 & 18 & 16 & 17 \\
CMS double diff. Drell Yan         &132 & 159  & 151  & 149 \\ 
\hline
    \end{tabular}
\end{center}
%\vspace{-.0cm}
\caption{\sf The quality of the description (in terms of $\chi^2$) of the LHC $W,Z$ data before and after they are included in the 
global NLO and NNLO fits. 
$\chi^2$ values obtained in 
the CPdeut fit of the NLO MMSTWW analysis \cite{MMSTWW} are shown for comparison. The data included were identical to those in MSTW08.}
%\label{tab:LHCWZ}
\label{table:MMHT14}
\end{table*}

%\subsubsection{NNLO PDF updates}
%\label{}

\subsection{Doubly differential Drell-Yan at CMS}
\label{}

\begin{figure}[t!]
  \label{fig:dyatcms}
  \centering
  \includegraphics[width=0.45\textwidth]{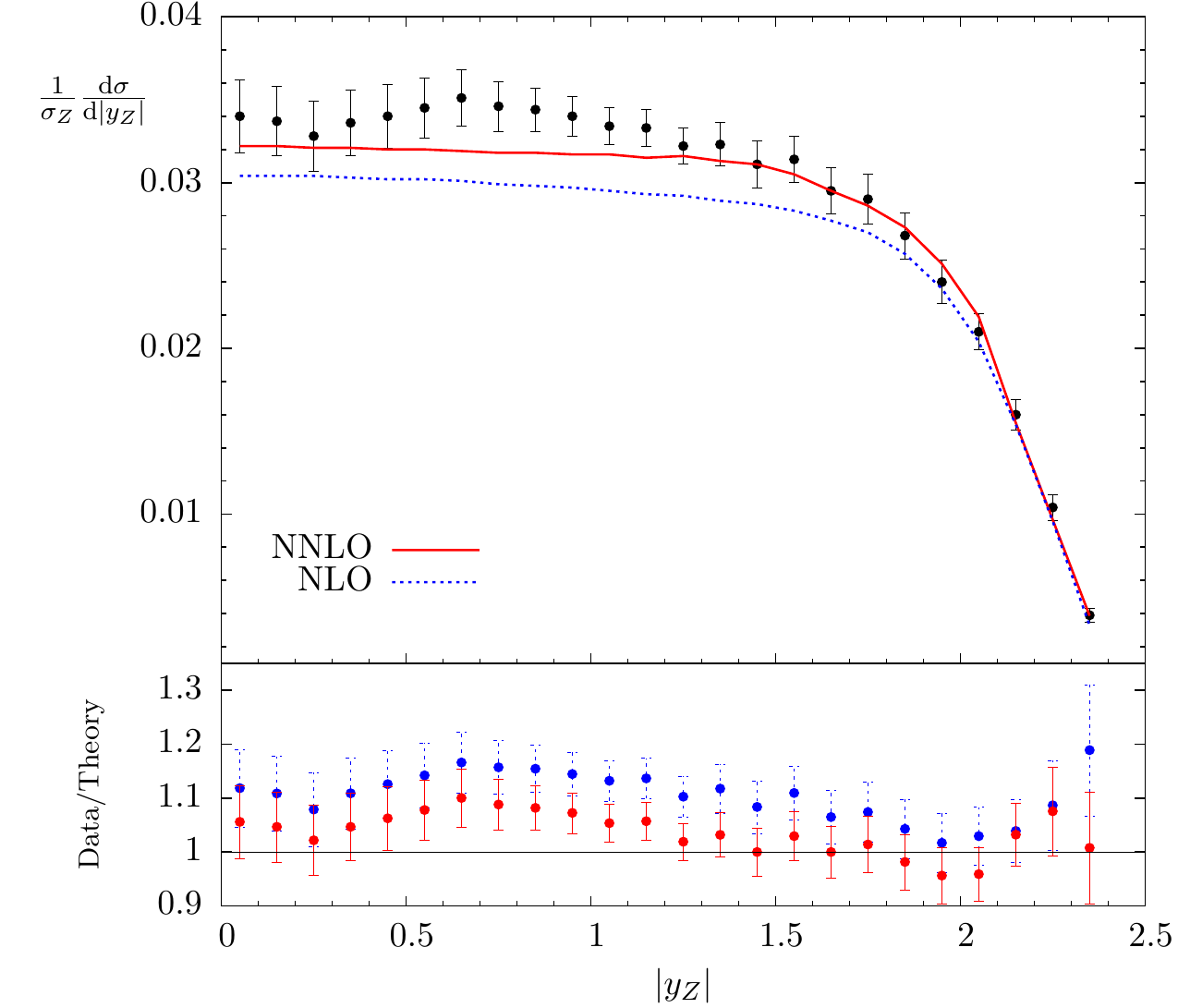}
  \includegraphics[width=0.45\textwidth]{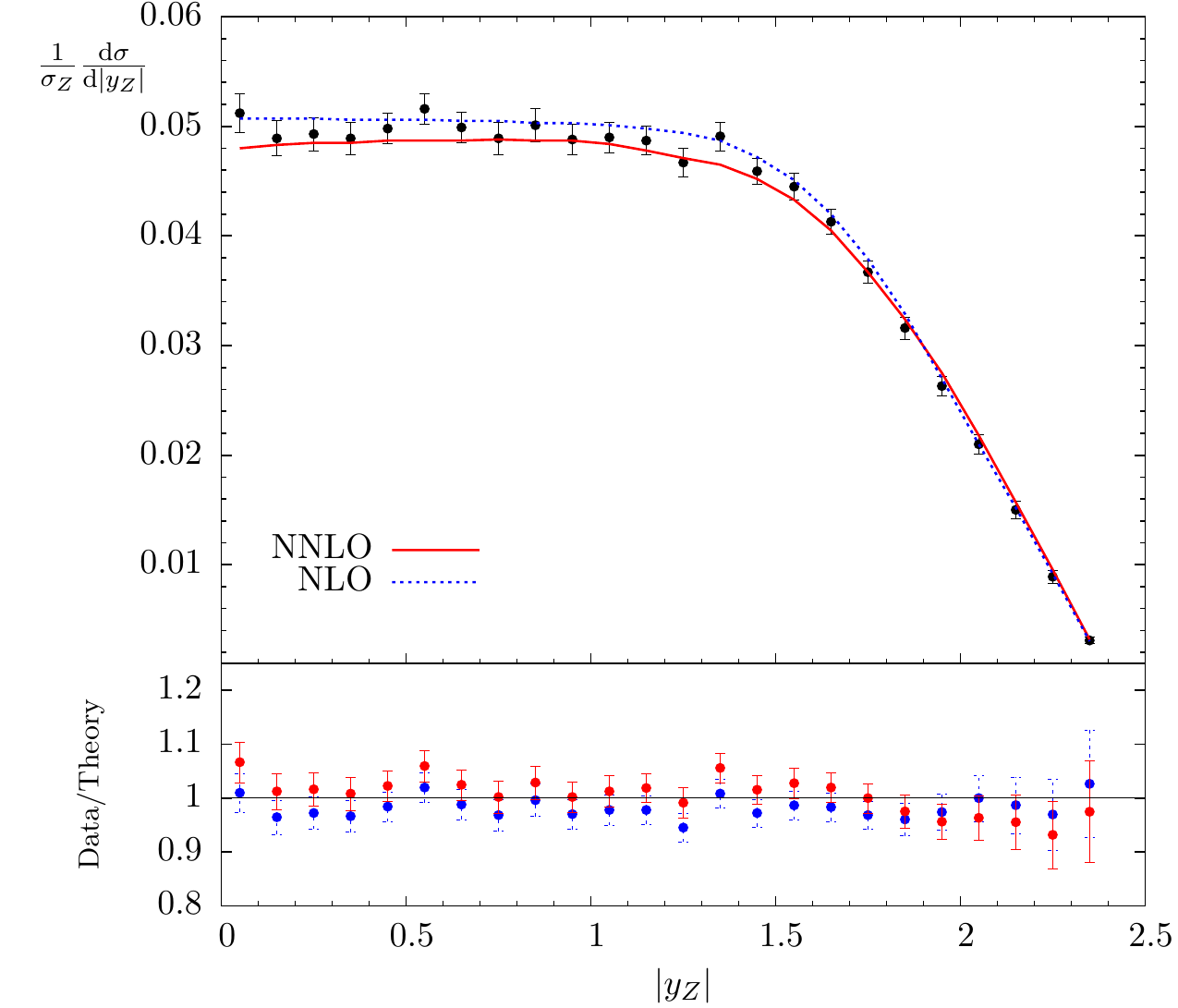}
  \caption{The two lowest mass bins for the doubly differential Drell-Yan cross section at CMS.}
\end{figure}

In the process of including LHC data into the PDF fits the doubly differential Drell-Yan data at CMS have been taken into account as well~\cite{CMS-ddDY}. The comparison between data and theoretical predictions at both NLO and NNLO are shown in Fig.~\ref{fig:dyatcms} for the two lowest mass bins. It seen that the NLO theory prediction describes the data rather poorly in the lowest mass bin, while NNLO is visibly closer to data. Things improve as we move higher up in mass. The LO component is very small in the lowest mass bin and what we see is the qualitative difference in description by the NLO and NNLO corrections in the low-mass limit.\\

\subsection{Results}
\label{sec:results}

In Figs.~\ref{fig:nloglu} and~\ref{fig:lightnlo} the impact of including LHC data on the gluon distribution and the light quark distributions, respectively, is shown at NLO. It is clear that the impact is limited in size with the most visible differences manifesting themselves at very low- and high $x$.\\
Figs.~\ref{fig:nnloglu} and~\ref{fig:strangennlo} show the impact of LHC data at NNLO on the gluon- and strange distributions, respectively. In the gluon case the change is rather moderate and, again, the uncertainty is lowered, generally. The strange distribution is seen to be affected significantly with the uncertainty becoming larger. This is predominantly due to the introduction of the charm to muon branching ratio of $B_{\mu}=0.092$ with the sizeable accompanying uncertainty of $\pm 10\%$. At NNLO we extract the value $\alpha_S=0.1172$.\\
The overall effect of including LHC data has been summarised in Table~\ref{table:MMHT14}. For NLO we see that the quality of the fit is improved by including LHC data and only in the case of the high-mass ATLAS Drell-Yan data are very slightly worse. In the NNLO case there is an overall improvement for all data, including the high-mass ATLAS Drell-Yan.

\section{Conclusions}
\label{}

In the past years there has been ongoing work on updating the PDFs and an official PDF update is to be released very soon.\\
In the process we have seen the inclusion of the most up-to-date HERA and Tevatron data. Furthermore the most relevant published LHC data have been included as well: ATLAS, CMS and LHCb $W,Z$--data, $t\bar{t}$--cross section data and all published ATLAS and CMS inclusive jet data, though the latter are currently not included at NNLO. \\
Inclusion of the mentioned data has so far had very few dramatic effects on the PDFs. The largest effect has been on the strange quark distribution (and mainly at NNLO) and low--$x$ valence quarks, the latter largely due to change in methodology and newer data. \\
Currently there is some uncertainty in the manner the available NNLO corrections may affect the fits to jet data. Due to this it has been decided to wait with the inclusion of LHC jet data at NNLO until the full fixed order NNLO calculations are finalised. 

\section*{Acknowledgements}
\label{acknowledge}

We would like to thank W. J. Stirling and G. Watt for numerous discussions on PDFs. This work is supported partly by the London Centre for Terauniverse Studies (LCTS), using funding from the European Research Council via the Advanced Investigator Grant 267352. RST would also like to thank the IPPP, Durham, for the award of a Research Associateship. We would like to thank the Science and Technology Facilities Council (STFC) for support.

%\section{}
%\label{}



%% References

%\nocite{*}
\bibliographystyle{elsarticle-num}
\bibliography{references}
%\bibliography{martin}

%\bibliography{references}{}
%\bibliographystyle{h-physrev}





%\input{proceedings.bbl}


% \end{thebibliography}

\end{document}